\documentstyle[12pt]{article}
\newcommand{\ben}{\begin{equation}}{\rm }
\newcommand{\een}{\end{equation}}
\newcommand{\bea}{\begin{eqnarray}}{\bf }
\newcommand{\eea}{\end{eqnarray}}
\newcommand{\bear}{\begin{array}}
\newcommand{\enar}{\end{array}}
\newcommand{\bdm}{\begin{displaymath}}
\newcommand{\edm}{\end{displaymath}}
\newcommand{\no}{\nonumber}
 
\newcommand{\hf}{\frac{1}{2}}

\newcommand{\La}{{\cal L}}

\newcommand{\pa}{\partial}
\newcommand{\del}{\delta}
\newcommand{\eps}{\epsilon}
\newcommand{\al}{\alpha}

{\bf }

\newcommand{\g}{\gamma}

\newcommand{\br}{\langle}
\newcommand{\kt}{\rangle}

\newcommand{\ps}{\psi}
\newcommand{\psb}{\overline{\psi}}

\newcommand{\half}{\frac{1}{2}}
\newcommand{\p}{\phi}

\newcommand{\be}{\beta}
\newcommand{\beb}{\bar{\beta}}
\newcommand{\gb}{\bar{\gamma}}
\newcommand{\r}{\rho}
\newcommand{\e}{\eta}
\newcommand{\pab}{\bar{\partial}}
\newcommand{\li}{\rightarrow}
\newcommand{\tps}{\tilde{\psi}}
\newcommand{\tbe}{\tilde{\beta}}
\newcommand{\sig}{\sigma}
\newcommand{\ra}{\rightarrow}
\newcommand{\fb}{\bar{f}}
\newcommand{\Gt}{\tilde{G}}
\newcommand{\la}{\lambda}
\newcommand{\chb}{\bar{\chi}}
\newcommand{\Gh}{\hat{G}}
\newcommand{\Gb}{\bar{G}}
\newcommand{\Ghb}{\bar{\hat{G}}}

\begin{document}
\def\ads3{$AdS_3$}
\def\alp{{\alpha}_+}
\parskip 5pt plus 1pc
\parindent=16pt

\begin{flushright}
AS-ITP-98-16 \\
\vspace{-1ex}
HEP-TH/9812216\\  
\vspace{-1ex} 
December, 1998
\end{flushright}
\vspace{2ex}
\vspace{1ex}
\vspace{1ex}
\vspace{16ex}
\centerline{\Huge Light-Cone  Gauge Quantization}
\vspace{1ex}
\centerline{\Huge of String Theories on $AdS_3$ Space}
\vspace{2ex}
\vspace{1ex}
\vspace{1ex}
\vspace{1ex}
\vspace{2ex}
\vspace{2ex}
\vspace{3ex}
\centerline{\large {\sc Ming Yu }{\footnote{\tt e-mail: yum@itp.ac.cn}}  
and {\sc Bo Zhang} {\footnote{\tt e-mail: zhangb@itp.ac.cn}}}
\vspace{2ex}
\centerline{\it Institute of Theoretical Physics, Academia Sinica}
\vspace{1ex} 
\centerline{\it P.O.Box 2735, Beijing 100080, P.R.China}    
\vspace{10ex}
\centerline{\large \it Abstract}
\vspace{1ex}
\begin{center}
\begin{minipage}{130mm}
{Light-cone gauge quantization procedures are  given, for superstring theory
on $AdS_3$ space charged with NS-NS background, both in NSR and GS formalism. 
The spacetime (super)conformal algebras are constructed in terms of the 
transversal physical degrees of freedom. The spacetime conformal anomaly 
agrees with that of covariant formalism, provided that the worldsheet conformal
anomaly $c$ equals 26 or 15 for bosonic string or superstring, 
respectively. The spacetime (super)conformal field theory is found to 
correspond to orbifold construction on symmetric product space 
$\it{Sym_p} {\cal{M}}/Z_p$.}
\end{minipage}
\end{center}
\vfill
\newpage

\section{Introduction}
Recently, String theories on $AdS^3$ space has attracted much attention 
\cite{sei, stro, ms, raj, pes, duf, kut, deb,  eli, par, sug}. 
In ref. \cite{mal}, Maldacena has  conjectured  that a type IIB 
superstring theory on $AdS_{d+1}$ is dual to a superconformal field theory 
living on the d-dimensional boundary of $AdS_{d+1}$, which is a compactified 
Minkovski space. A more precise statement on the $AdS/CFT$ correspondence is 
give in refs. \cite{wit1,pol}, which relate the boundary values in the former 
theory to the source terms in the later.
Some well studied examples of the $AdS/CFT$ correspondence are the cases 
of $AdS_5$ and $AdS_3$. On each boundary resp., one finds a 4d $N=4$ 
superconformal Yang-Mills theory in the former case, and  a 2d $N=4$ 
superconformal field theory in the later. In order to substantialize 
Maldacena's conjecture, a better understanding on the physical contents
of the SCFT living on the boundary of $AdS$ space is
required. While for the case of $AdS_5$, our knowledge is limited on 
computing
the physical correlators in $N=4$ superconformal Yang-Mills theory,
the case of $AdS_3$ space should provide a better test ground, thanks to the
conformal Ward identities of 2d nature, which involves infinitely many
conserved charges. Hopefully, the techniques developed in solving the
$AdS_3/CFT$ correspondence can be eventually generalized to the cases of higher
dimensional $AdS$ spaces.  

Presently, there are two main approaches to formulate
string theories on $AdS_3$ space, analogous 
to the development of the  superstring theories by the NSR and 
GS formalism.   

(1) As for the NSR formalism, a quantization procedure on $AdS_3$ space is 
given in 
ref.\cite{sei}, in which, the spacetime superconformal generators are 
constructed and some physical states given. 
The advantage of the NSR formalism is its manifest covariance and 
worldsheet superconformal invariance. The disadvantage is that the spacetime 
supersymmetry  is not manifest and it is difficult to couple it to the R-R 
charged background. Another disadvantage is that the physical state content
is not manifest and deserves a careful BRST cohomology analysis.

(2) For the case of Green-Schwarz type superstring on $AdS_3$ space, a 
$\kappa$ symmetry invariant action and its simplified 
($\kappa$ symmetry fixed) 
version are given in refs.\cite{raj,pes}.\footnote{The nonlinear $\sigma$ 
model on coset space $\frac{SU(2,2|4)}{SO(4,1)\times SO(5)}$
are first proposed to describe superstring
theory on $AdS_5$ space. For more details, see refs.\cite{tse}. The \ads3
case is a generalization from that in ref.\cite{tse}.}
 In the Green-Schwarz formalism, 
the spacetime supersymmetry is manifest, and both NS-NS and R-R background 
can be charged. However, due to its high non-linearality,
the quantization procedure for the present case is not 
known, even for its simplified version. 
Therefore, both  worldsheet and spacetime superconformal invariance are not 
obvious at the quantum level.

The purpose of the present paper is trying to make a connection between
the two independent approaches. Our starting point is the NSR formalism.
After fixing the conformal gauge, there are still residual conformal 
gauge symmetries, which are responsible for getting rid of negative norm states
and make the string theory unitary. The usual covariant quantization is to 
enlarge the Hilbert space to include the reparametrization ghosts, and the 
negative norm states decouple from the physical state space which are BRST 
invariant. The underlying spacetime field theory should only depend on the
physical degrees of freedom, which are transversal to the longitudinal ones. 
However, there are no prescribed ``longitudinal'' and ``transversal'' 
directions in the Hilbert space, each choice should give 
the equivalent string theory as ensured by the conformal invariance.

In contrast, in light-cone gauge quantization, the 
conformal gauge symmetry is completely fixed, and one has to make a definite 
choice on
which are the longitudinal degrees of freedom and which are the transversal
ones. In flat spacetime, the light-cone gauge choice is made to align 
worldsheet ``time'' direction along a light-cone coordinate in the spacetime. 
In our $AdS_3$ case things are more complicated, since, generally 
in curved spacetime there can be no flat directions along which light 
moves freely.

In the present paper, we shall propose
a way to resolve such kind of problems. The point is that at the $AdS_3$
boundary, one can find a flat light-cone direction, and at the vicinity of
the boundary, the term which curve the space-time is really the screening 
charge, which can be perturbatively expanded as what are usually done in the 
calculations of the correlators in WZNW models, cf. ref.\cite{pry}. 

To justify our approach to the quantizations
of string theories on $AdS_3$ space, we have checked the
following consistency conditions:

(1) The spacetime superconformal algebra proposed in ref.\cite{sei},
when re-expressed in our transversal degrees of freedom after the
light-cone gauge choice is made, is closed. The conformal anomaly
agrees with that given in ref.\cite{sei}, provided the worldsheet
conformal anomalies $c$ equal 26 resp. 15 for bosonic string resp. 
superstring.

(2) In string theories on $AdS_3$ space,  there is a flat time direction
which is dual to the $J^3$ in $sl(2,R)$ current algebra. 
Our light-cone gauge choice is the same as choosing this flat time
direction combined with another flat space direction to form a light-cone.
Since there exist unitary representations in $SL(2,R)/U(1)$ conformal
field theory, cf. ref.\cite{dix}, the unitarity of the string theory is 
ensured, provided the non-$AdS_3$ part is made of unitary conformal field 
theories.
 
 Using the light-cone gauge quantization to get rid of the unphysical 
degrees of freedom, we recover the spacetime superconformal algebra in terms 
of the
physical transversal degrees of freedom. By this method, we find that the
spacetime superconformal field theory is described by the orbifold theory 
on the target space 
$\frac{Sym_p({\mathcal M}_{Liouville} \times S_3 \times T^4)}{Z^p}$,
and spacetime supersymmetry is realized in a nontrivial way. By the similar 
procedure, the GS superstring theory on $AdS_3 \times S_3
\times T_4 $ with NS-NS background \cite{raj,pes} can be quantized in 
the light-cone gauge, giving the same spacetime superconformal algebra 
as the NSR superstring theory. The generalization 
of our procedure to the case of R-R charged background is also currently under 
investigation.\cite{later} 

The present paper is organized in the following way. In section 2, we review 
the NSR formalism of the bosonic string theory on $AdS_3 \times \cal{N} $,
and give a description of how to make the light-cone gauge choice and what the
spacetime conformal field theory is. 

In section 3, we generalize the light-cone gauge quantization procedure to the
case of fermionic string. A consistent quantization procedure, which makes 
light-cone gauge choice possible, is given.

In section 4, the explicit construction of the spacetime $N=4$ $su(2)$ 
extended superconformal algebra is given in terms of fields living in 
transverse spacetime.

In  section 5, we quantize the Green-Schwarz type superstring theory on 
$AdS_3 \times S_3 \times T_4 $ spacetime with NS-NS background.

In section Conclusions and Discussions, we speculate on how to calculate 
one loop partition functions and what could be further done using our 
generalized light-cone gauge quantization procedure. 
  
Appendix A is devoted to clarify our conventions, in particular for the spinors
in \ads3 space, as well as the relation between the $N=2$ spacetime SUSY 
algebra 
$su(1,1|2)^2$ and the global part of the $N=4$ spacetime SCA.

\section{Bosonic String Theory on $AdS_3$ Space}
\subsection{String Action and Light-Cone Gauge Quantization}
$AdS_3$ space is a hypersurface embedded in flat ${\mathcal R}^{2,2}$,        
\ben 
-X_{-1}^2+X_1^2+X_2^2-X_3^2=-l^2 
\een
with the induced metric
\ben
ds^2=-dX_{-1}^2+dX_1^2+dX_2^2-dX_3^2
\een
It admits the  $sl(2,R) \times sl(2,R)$ Lie algebra as its isometry generators,
\ben
[J_{\al\be},J_{\g\r}]={\e}_{\al\r}J_{\be\g}+{\e}_{\be\g}J_{\al\r}
-{\e}_{\be\r}J_{\al\g}-{\e}_{\al\g}J_{\be\r}\label{cmj}
\een
where ${\e}_{\al\be}$ is the flat metric $diag(-1, 1, 1, -1)$.
Let
\bea
J^1&=&-i\half(J_{23}+J_{-11}) \no\\
J^2&=&-i\half(J_{31}+J_{-12}) \no \\
J^3&=&-i\half(J_{12}-J_{-13}) \no  \\
\bar{J}^1&=&-i\half(J_{23}-J_{-11}) \\
\bar{J}^2&=&-i\half(J_{31}-J_{-12}) \no \\
\bar{J}^3&=&-i\half(J_{12}+J_{-13}) \no 
\eea
The Commutators eq.(\ref{cmj}) become now
\bea
[J^A , J^B]&=&i{\e}_{CD}{\eps}^{ABC}J^D \no \\\
[\bar{J}^A ,\bar{J}^B]&=&i{\e}_{CD}{\eps}^{ABC}\bar{J}^D  \\\
[J^A, \bar{J}^B]&=&0 \no  
\eea
In conformal field theory, the $J^A$'s resp. $\bar{J}^A$'s are just the zero 
modes of the affine 
$sl(2,R)\times sl(2,R)$ Lie algebra at level k generated by worldsheet 
holomorphic currents $J^A(z)$ resp. anti-holomorphic ones 
$\bar{J}^A(\bar{z})$, satisfying  the OPE:
\ben
J^A(z)J^B(w)=\frac{k{\e}^{AB}/2}{(z-w)^2}+\frac{i{\e}_{CD}{\eps}^{ABC}
J^D}{z-w}+\cdots\cdots \label{ope}
\een
where A,B,C,D=1,2,3 and ${\e}_{AB}=diag(+ + -)$, and similar OPE for the
anti-holomorphic ones. 
\\

It is convenient to choose the independent coordinates on \ads3 space as
\bea
\g &=& \frac{X_2+iX_1}{X_{-1}+X_3} \\
\bar{\g}&=& \frac{X_2-iX_1}{X_{-1}+X_3}\no\\
\p &=& \frac{log(X_{-1}+X_3)}{l}\no
\eea

Then the metric tensor becomes
\ben
ds^2=l^2(d\p^2 + e^{2\p}d\g d\bar{\g})
\een
As given in ref.\cite{sei}, the Lagrangian for the bosonic string theory
on ${AdS_3}\times \cal{N}$ with NS-NS background is
\ben
\La=\pa\p\pab\p-\frac{2}{\alp}{\hat{R}}^{(2)}\p+\be\pab\g
+\beb\pa\gb-\be\beb exp(-\frac{2}{\alp}\p)+{\La}_N \label{la}
\een
where ${\alp}^2=2k-4$, $k=\frac{l^2}{l_s^2}$, ${\La}_N$ is the contribution
from manifold $\cal{N}$.
The conformal weights of $\be$,$\g$ are 1, 0 respectively.\\
The conformal invariant term $\int \be\bar{\be} exp(-\frac{2\p}{\alp})$
in the action
represents the screening charges in the calculations of the correlators
in the WZNW theory. Treating it as a perturbation term, then the classical 
equations of motion for $\g , \be ,\p $ become free ones. The OPE of these 
fields are
\bea
\p(z)\p(0)&=&-log(z)+\cdots\cdots\\
\be(z)\g(0)&=&\frac{1}{z}+\cdots\cdots \no
\eea
and others are regular.

The $sl(2,R)$ current algebra can be represented by 
\bea
J^3&=&\be\g+\frac{\alp}{2}\pa\p \no \\
J^+&=&\be{\g}^2+\alp\g\pa\p+k\pa\g \label{rep}\\
J^-&=&\be \no
\eea
Here, $J^+=J^1+iJ^2$, $J^-=J^1-iJ^2$.

The energy-momentum tensor is given by Sugavara construction:
\ben
T=-\half\pa\p\pa\p+\be\pa\g-\frac{1}{\alp}{\pa}^2\p+T_N\label{ttt}
\een
where $T_N$ is the energy-momentum tensor coming from the manifold $\cal{N}$.

When $\frac{\p}{\alp}$ is large, we can perturbatively expand the term 
$e^{\int \beb \be exp(-\frac{2\p}{\alp})}$
as screening charges. That means the  string is really located in the 
neighborhood of $AdS_3$ boundary $(\frac{2\p}{\alp}\li\infty)$. Module
possible contractions with screening charges in the correlators,
$\g $ can be treated as a holomorphic field with conformal weight 0. 
By  conformal invariance, we can transform $\g $ into 
a desired form containing no string excitations.
It is convenient to make our light-cone gauge choice as follow:
\ben
\g=e^qz^p \label{gamma}
\een
Here, the momentum $p=\oint\g^{-1}(z)\pa\g(z)$ equals the number of times
the string winds in the $\g$ space, thus must be an integer. And $e^q$ 
represents the center of mass coordinate of the string, is not an important
factor, since it can always be scaled away by a dilation in $z$, 
$z\ra e^{q/p}z$.
If the context is clear, we can always set $q=0$. However, keeping the factor
$e^q$ can remind us that $\g$ is in fact a worldsheet scalar, since $e^q$ adds
conformal weight $p$ to the vacuum.  

Fixing $\g$ we can solve the  constraint $T=0$, where $T$ is defined in 
eq.(\ref{ttt}), for $\beta$:
\ben
\be(z)=\frac{1}{p}e^{-q}z^{1-p}(\half\pa\p(z)\pa\p(z)
+\frac{1}{\alp}{\pa}^2\p(z)-T_N(z)
+\frac{\Delta}{z^2})\label{beta}
\een
where $\Delta$ is a normal ordering constant.

To justify this as the right light-cone gauge choice and that the 
transversal degrees of freedom is ghost free, let us first bosonize the 
$\be$, $\g$ system.
\bea
\g(z) &=& e^{\p_1(z)+\p_2(z)} \no \\
\be(z) &=& -\pa \p_1(z) e^{-\p_1(z)-\p_2(z)} \no \\
\p(z) &\equiv & {\p}_3(z)\\
\p_i(z)\p_j(w)&=& \eta_{ij}\ log(z-w)\no \\
\eta_{ij}&=& Diag\{1,-1,-1\} \no
\eea

$J^3(z)$ and the stress tensor can be expressed in terms of $\p_i(z)$ field as
\bea
J^3(z)&=&-\pa\p_2(z)+\hf \alp \pa\p_3(z) \\
T(z)&=&\hf \eta^{ij} \pa \p_i(z) \pa \p_j(z)-\hf\pa^2[\p_1(z)+\p_2(z)+
\frac{2}{\alp}\p_3(z)]
\eea

It is convenient to make an $O(1,2)$ transformation on the scalar fields,
\ben
\left(\begin{array}{c}
\Phi_1(z)\\ \Phi_2(z)\\ \Phi_3(z)\end{array}\right)
=\left( \begin{array}{lll}
\frac{\sqrt{4+\alp^2}}{2} & \frac{\alp^2}{2\sqrt{4+\alp^2}} & 
\frac{\alp}{\sqrt{4+\al^2}} \\
0 & \frac{2}{\sqrt{4+\alp^2}} & 
\frac{-\alp}{\sqrt{4+\alp^2}} \\
\hf\alp & \hf \alp & 1
\end{array} \right)
\left( \begin{array}{c} \p_1(z)\\ \p_2(z) \\ \p_3(z) \end{array} \right)
\label{trp}
\een

which can be solved to give
\ben
\left(\begin{array}{c}
\p_1(z)\\ \p_2(z)\\ \p_3(z)\end{array} \right)
=\left( \begin{array}{lll}
\frac{\sqrt{4+\alp^2}}{2} & 0 & -\hf\alp\\
\frac{-\alp^2}{2\sqrt{4+\alp^2}} & \frac{2}{\sqrt{4+\alp^2}} &\hf\alp\\
\frac{-\alp}{\sqrt{4+\al^2}} &\frac{-\alp}{\sqrt{4+\alp^2}} & 1
\end{array} \right)
\left( \begin{array}{c} \Phi_1(z)\\ \Phi_2(z) \\ \Phi_3(z) \end{array} \right)
\een
 
In terms of $\Phi_i(z)$ fields we have
\bea
J^3(z)&=&-\sqrt{\frac{k}{2}}\pa \Phi_2(z) \\
T(z)&=&\hf \eta^{ij} \pa \Phi_i(z) \pa \Phi_j(z) 
-\frac{1}{\alp}{\pa}^2\Phi_3(z)+T_N(z) \label{tensor}\\
\g(z)&=&e^{\sqrt{\frac{2}{k}}(\Phi_1(z)+\Phi_2(z))} \\
\be(z)&=& (\sqrt{\frac{k}{2}}\pa \Phi_1(z)-\hf\alp\pa \Phi_3(z))
e^{\sqrt{\frac{2}{k}}(-\Phi_1(z)-\Phi_2(z))}
\eea

And the screening charge becomes now.
\ben
V_+ \equiv\oint dz \be (z)e^{-\frac{2}{\alp}\p (z)} 
=\oint dz \sqrt{\frac{k}{2}}\pa \Phi_1(z)
e^{-\sqrt{\frac{2}{\alp}}\Phi_3(z)}
\een
where we have dropped a total derivative term.

From the stress tensor eq.(\ref{tensor}), we see that there are two flat
spacetime directions, $\Phi_1$ and $\Phi_2$, the former is space-like,
and the later, which is dual to the current $J^3$, is time-like.
Using the residual conformal invariance in the conformal gauge,
we can transform the worldsheet ``time'' direction parallel to the coordinate
$\Phi^+(z,\bar z)\equiv \Phi_1(z,\bar z)+\Phi_2(z,\bar z)$, i.e.
\ben
\Phi^+(z,\bar{z})=2q^++2p^+\tau
\een

From eq.(\ref{trp}) we have $\p_1+\p_2 \propto \Phi_1+\Phi_2$. We find that 
making $\Phi_1+\Phi_2$ the light-cone coordinate is equivalent to making 
$\g (z)$ in the form given by eq.(\ref{gamma}), where $\p^+=\p_1+\p_2$ is
fixed to be $\p^+=2q+2p\tau$. It is well known that there
exist unitary representations in
$SL(2,R)/U(1)$ coset space theory with $U(1)$ corresponding to the time-like 
direction \cite{dix}. We conclude that the transversal physical space
is ghost free provided the conformal field theory from $\cal{N}$
is unitary.

\subsection{Spacetime Virasoro Algebra in Light-Cone Gauge}

At the \ads3 boundary, $\p \ra \infty$, the action given by eq.(\ref{la}) is 
invariant under spacetime conformal transformations:
\bea 
\g &\ra& f(\g) \\
\gb &\ra& \bar{f}(\gb) \\
\be &\ra& \be f'(\g)^{-1} + \pa \p f''(\g)^{-1} \\
\beb &\ra& \beb \fb'(\gb)^{-1} + \pab \p \fb''(\gb)^{-1} \\
\p &\ra& \p -\frac{\alp}{2}log(f'(\g)\fb'(\gb))
\eea

The Virasoro algebra, generated by the conformal transformations on the 
boundary of $AdS_3$ space, is constructed in ref.\cite{sei} as
\ben
L_n=-\oint dz[(1-n^2)J^3{\g}^n-\frac{n(n-1)}{2}J^-{\g}^{n+1}
+\frac{n(n+1)}{2}J^+{\g}^{n-1}]\label{cln}
\een

The spacetime conformal transformations act on the physical space of the 
string theory, which are invariant under worldsheet conformal transformations.
Thus, it is possible to express the spacetime Virasoro generators in terms of 
fields in transversal dimensions. Substituting our light-cone gauge choice, 
eqs.(\ref{gamma}, \ref{beta}) into eq.(\ref{cln}) for the spacetime Virasoro
generators, we have
\bea
L_n&=&\oint dz[\frac{1}{p}(-\half\pa\p\pa\p-\frac{1}{\alp}{\pa}^2\p+T_N
-\frac{\Delta}{z^2})
e^{nq}z^{np+1}-\frac{\alp}{2}(n+1)\pa\p e^{nq}z^{np}]\label{lln}\\
&=&\oint dz\lbrace\frac{1}{p}[-\half\pa\p\pa\p-(\frac{1}{\alp}-\frac{\alp}{2})
{\pa}^2\p+T_N-\frac{\Delta}{z^2}]e^{nq}z^{np+1}
-\frac{\alp(p-1)}{2p}\pa\p e^{nq}z^{np}\rbrace \no
\eea

Using the OPE of $T_N$ and $\p$ , it is ready to check that that the $L_n$'s 
in the light-cone gauge constructed above close a Virasoro algebra 
\ben
[L_n, L_m]=n(n^2-1)\frac{pc}{12}{\del}_{m+n, 0}+(n-m)[L_{n+m}+
\del_{n+m,0}(\frac{a}{2}+\frac{\Delta}{p})]\label{lnlm}
\een
with
\bea
c&=&c_N+1+12(\frac{1}{\alp}-\frac{\alp}{2})^2\\
a&=&\frac{c_N+1+\frac{12}{{\alp}^2}}{12}(p-\frac{1}{p})-(p-1) 
\eea
where $c_N$ is the central charge of $T_N$.\\

If we shift $L_0$ by $\frac{a}{2}+\frac{\Delta}{p}$, 
eq.(\ref{lnlm}) becomes the standard form of 
the Virasoro algebra with central charge $cp$. 

In covariant quantization, the central charge of this
spacetime Virasoro algebra is $6kp$, so to get rid of spacetime anomaly 
we must set $cp=6kp$. This condition fixes the worldsheet center charge for 
the string theory to be 
\ben
c_{string}=c_N+3+12\frac{1}{{\alp}^2}=26
\een
where ${\alp}^2=2k-4$.

That means 
when the string propagates in critical dimension, its light-cone 
gauge quantization could be equivalent to the covariant one. 
In critical dimension, the light-cone gauge fixed Virasoro generators should
close the same algebra as the covariant one, i.e. the shift of $L_0$ should 
be $0$, or $\Delta=-\frac{pa}{2}$.
Then the Virasoro generators of spacetime CFT can be 
written as
\ben
L_n=\frac{1}{p}\tilde L_{np}+\frac{c}{24}(p-\frac{1}{p})\delta_{n,0}\label{dlln}
\een
where
\ben
\tilde L_m=\oint dz[T_N-\half\pa\p\pa\p-(\frac{1}{\alp}-\frac{\alp}{2}){\pa}^2\p]
z^{m+1}
\een
We shall show below that eq.(\ref{dlln}) defines a conformal transformation: 
$z\ra w=z^{\frac{1}{p}}$
\bea
T(z)&=&(w'(z))^2T(w)+\frac{c}{12}[\frac{w'''(z)}{w'(z)}
-\frac{3}{2}(\frac{w''(z)}{w'(z)})^2]\\
&=&\frac{1}{p^2}z^{\frac{2}{p}-2}T(w)+\frac{c}{24}(1-\frac{1}{p^2})z^{-2}\no
\eea
Only the term with integer power of $z$ are well defined on the $z$ complex
plane closing a Virasoro Algebra. Hence
\bea
L_n&=&\int dz T(z)z^{n+1}\label{dlln2}\\
&=&\frac{1}{p}\int dw T(w)w^{pn+1}+\frac{c}{24}(p-\frac{1}{p})\del_{n,0}\no\\
&=&\frac{1}{p}\tilde L_{pn}+\frac{c}{24}(p-\frac{1}{p})\del_{n,0}\no
\eea
which agrees with eq.(\ref{dlln}).

In fact, eq.(\ref{dlln}) suggests that the spacetime CFT is an orbifold theory 
on the target space $\frac{Sym_p{\mathcal M}}{Z^p}$,
see subsection \ref{orb} for detailed discussions.

\section{Light-Cone Gauge Quantization of Fermionic String Theory on $AdS_3$}

The light-cone gauge quantization for bosonic string theory on \ads3 space 
discussed in the previous section is ready to be generalized to the fermionic 
case.
For fermionic string theory on the group manifold, the worldsheet 
supersymmetry is achieved by considering super WZNW theory \cite{dkpr}. 
For each
bosonic current, we associate a fermionic partner, which
is a worldsheet real fermion field and living in the adjoint representation of
the finite dimensional Lie algebra.

For fermionic string theory on \ads3 space, the affine $sl(2,R)$ currents 
pertaining to the  $AdS_3$ space can be written as
\ben
J^A=j^A-i\frac{1}{k}{\eps}^A_{BC}{\ps}^B{\ps}^C\\
\een
which satisfy OPE eq.(\ref{ope}).
Here we use the convention as in ref. \cite{sei}
\ben
\br{\ps}^A(z){\ps}^B(w)\kt=\frac{k{\e}^{AB}/2}{z-w}
\een
The worldsheet supercurrent $G$ and the energy-momentum tensor are
\bea
G(z)&=&\frac{2}{k}({\e}_{AB}{\ps}^Aj^B
-\frac{i}{3k}{\eps}_{ABC}{\ps}^A{\ps}^B{\ps}^C)+G_N\\
T(z)&=&\frac{1}{k}(j^Aj_A-{\ps}^A\pa{\ps}_A)+T_N \no\label{gt}
\eea 
Set 
\bea
J^+&=&J^1+iJ^2	\hskip 3cm	J^-=J^1-iJ^2 \no \\
j^+&=&j^1+ij^2	\hskip 3cm	j^-=j^1-ij^2\\
{\ps}^+&=&{\ps}^1+i{\ps}^2	\hskip 3cm {\ps}^-={\ps}^1-i{\ps}^2 \no
\eea
$j^A$'s can be represented as in eq.(\ref{rep}), but now with ${\alp}^2=2k$.

For light-cone gauge quantization, it is convenient to use another set of 
free fermion fields
\bea
{\tps}^3&=&{\ps}^3-{\ps}^-\g \no \\
{\tps}^+&=&{\ps}^++{\ps}^-{\g}^2-2{\ps}^3\g\\
{\tps}^-&=&{\ps}^- \no
\eea
And define 
\bea
\tbe&=&\be-\frac{2}{k}{\ps}^-{\ps}^3\\
&=&\be-\frac{2}{k}{\tps}^-{\tps}^3 \no
\eea
We then have
\bea
\br{\tps}^3(z){\tps}^3(w)\kt&=&\br{\ps}^3(z){\ps}^3(w)\kt=\frac{-k/2}{z-w}\\
\br{\tps}^+(z){\tps}^-(w)\kt&=&\br{\ps}^+(z){\ps}^-(w)\kt=\frac{k}{z-w} \no
\eea
and
\bea
\br\tbe(z){\tps}^A(w)\kt&=&0\\
\br\tbe(z)\g(w)\kt&=&\frac{1}{z-w} \no
\eea

Using these fields, the supercurrent and the energy-momentum tensor for the
worldsheet CFT can be written as
\bea
G&=&\frac{2}{k}(\half{\tps}^+\tbe-\frac{\alp}{2}{\tps}^3\pa\p
+\frac{k+2}{2}{\tps}^-\pa\g)+G_N\\
T&=&(\tbe\pa\g-\half\pa\p\pa\p-\frac{1}{\alp}{\pa}^2\p)-\frac{1}{k}(
-{\tps}^3\pa{\tps}^3+\half{\tps}^+\pa{\tps}^-+\half{\tps}^-\pa{\tps}^+)+T_N \no
\eea

The spacetime Virasoro algebra for fermionic string theory on \ads3 space is
defined in  ref.\cite{sei}
\bea
L_n&=&-\oint dz\oint dw G(z)[(1-n^2){\ps}^3{\g}^n
+\frac{n(n-1)}{2}{\ps}^-{\g}^{n+1}+\frac{n(n+1)}{2}{\ps}^
+{\g}^{n-1}]_{(w)}\label{cfln}\\
&=&-\oint dw[(1-n^2)J^3{\g}^n+\frac{n(n-1)}{2}J^-{\g}^{n+1}
+\frac{n(n+1)}{2}J^+{\g}^{n-1}]_{(w)} \no
\eea
Here
\bea
J^3&=&j^3+\frac{{\ps}^+{\ps}^-}{k}=(\tbe\g+\frac{\alp}{2}\pa\p)
	+\frac{{\tps}^+{\tps}^-}{k} \no\\
J^+&=&j^++\frac{2{\ps}^+{\ps}^3}{k}=(\tbe{\g}^2+\alp\g\pa\p+k\pa\g)
	+\frac{2}{k}{\tps}^+({\tps}^-\g+{\tps}^3)\\
J^-&=&j^--\frac{2{\ps}^-{\ps}^3}{k}=\tbe \no
\eea
In light-cone gauge quantization, similar to what  we have 
done in the bosonic case, we set 
\ben
\g=e^qz^p\label{gamma2}
\een
Making our gauge choice worldsheet supersymmetry 
invariant requires the OPE between $G$ and $\g$ be regular, so at the same 
time we set 
\ben
\tps^+=0 
\een

Now we can solve the constraints $G=0, T=0$, where $G,\ T$ are defined in 
eq.(\ref{gt}), for ${\tps}^-$ and $\tbe$.
\bea
{\tps}^-&=&\frac{e^{-q}z^{1-p}}{p(k+2)}[-kG_N+\alp{\tps}^3\pa\p]\no\\
\tbe&=&\frac{e^{-q}z^{1-p}}{p}[\half\pa\p\pa\p+\frac{1}{\alp}{\pa}^2\p
-\frac{1}{k}{\tps}^3\pa{\tps}^3-T_N+\frac{\Delta}{z^2}]\label{tbeta}
\label{beta2}
\eea

Substituting eq.(\ref{tbeta}) into eq.(\ref{cfln}), the spacetime Virasoro 
generators become
\ben
L_n=\oint dz [\frac{1}{p}(-\half \pa\p\pa\p- \frac{1}{\alp}{\pa}^2\p+T_M
	+\frac{1}{k}{\tps}^3\pa{\tps}^3-\frac{\Delta}{z^2})e^{nq}z^{np+1}
	-\frac{\alp}{2}(n+1)\pa\p e^{nq}z^{np}]\label{lfln}
\een
Comparing eq.(\ref{lfln}) for the fermionic string theory with eq.(\ref{lln}) 
for the bosonic case,  we find one  more term containing ${\tps}^3$ coming 
from \ads3 part, and 
${\alp}^2=2k$ now.

Similar to the bosonic case, if we require the $L_n$ defined in eq.(\ref{lfln})
close the same spacetime algebra as the covariant one with central charge 
$c=6kp$, then we must have
\bea
c_{string}&=&c_N+3+12(\frac{1}{{\alp}^2})+\frac{3}{2}=15\\
\Delta&=&-\hf[\frac{c_N+1+\frac{12}{\alp^2}+\hf}{12}(p^2-1)-(p^2-p)]
\eea
That agrees with the fact that the critical dimension of Fermionic 
string is 10.\\

\section{$N=4$ Spacetime SCFT}

For the superstring theory on \ads3 space, because of spacetime supersymmetry,
the CFT on the \ads3 boundary is in fact a superconformal field theory (SCFT).
For definiteness, we consider the spacetime manifold to be 
$AdS_3 \times S^3 \times T^4$ as an example. The spacetime $N=2$ supersymmetry
algebra coming from  $AdS_3 \times S^3$ is the global part of the $N=4$ 
su(2) extended superconformal algebra \cite{yu,ademollo} for the SCFT on the 
\ads3 boundary. In the following subsections, we shall first construct 
the $N=4$ SCA explicitly in term of physical transversal degrees of freedom,
and then find out the target space in which a possible $N=4$ nonlinear 
$\sigma$ 
models may live.

\subsection{Construction of $N=4$ Superconformal Currents}\label{4.1}
The affine su(2) algebra in the spacetime $N=4$ SCFT is lifted from the
worldsheet su(2) currents pertaining to the $S_3$ group manifold. 
For SU(2) super WZNW theory\cite{dkpr}, the worldsheet current algebra can be 
represented as
\bea
K^A&=&k^A+\half {\ps}_{\al} {\sig}^A_{\al \be} {\psb}_{\be}\\
k^A(z)k^B(w)&=&\frac{(k'-2)/2~{\del}^{AB}}{(z-w)^2}
+\frac{i{\eps}^{ABC}k^C}{z-w}+\cdots \cdots\\
K^A(z)K^B(w)&=&\frac{k'/2~{\del}^{AB}}{(z-w)^2}+\frac{i{\eps}^{ABC}K^C}{z-w}+
\cdots \cdots\no
\eea
Here, $k^A$'s are orthogonal to the $\ps^A$ fields. The zero modes
of the affine $su(2)$ currents $K^A(z)$'s represent one factor of 
the $SO(4)$ symmetry on the manifold $S^3$.
The spacetime affine $su(2)$ algebra is defined in ref.\cite{sei}
\ben
K^A_n=\oint K^A(z)\g^n dz
\een
In the light-cone gauge, it becomes simply
\ben
K^A_n=\oint K^A(z)e^{nq}z^{pn} dz=e^{nq}\tilde K^A_{np}\label{kn}
\een
It is clear that $K^A_n$'s form a affine $su(2)$ Lie algebra with level
$k_{space}=pk'$.
 
In eqs.(\ref{lfln},\ref{kn}), we have constructed the spacetime Virasoro 
generators $L_n$ 
and the affine $su(2)$ Lie algebra in the light-cone gauge. In order the 
spacetime
$N=4$ SCA close, we still need to construct the spacetime supercurrents in
the light-cone gauge. In ref.\cite{sei}, the global part of the $N=4$ 
spacetime superconformal 
algebra is constructed by covariant quantization, which involves worldsheet
superconformal ghosts and picture changing techniques. The full $N=4$ 
superconformal algebra can be generated by commuting $G^\al_r$'s with 
either $L_n$ or the $J^a_n$'s. However, such construction  can not be 
converted  into constructing the supercurrents in the light-cone gauge.  
The reason is in the later case, the spacetime supercurrents should involve
only the physical degrees of freedom, into which we do not know yet how to 
transform the worldsheet ghost part.      

Now it is clear that if our light-cone gauge choice, 
eqs.(\ref{gamma2}-\ref{beta2}), works, 
we should be able to construct the spacetime supercurrents in the
$N=4$ SCFT in terms of the transversal degrees of freedom. 
The superconformal algebra should close with the Virasoro generators and
the affine su(2) algebra given by eqs.(\ref{lfln},\ref{kn}). It turns out 
that indeed 
there exists a unique expression for the spacetime supercurrents and now we 
proceed to  construct them. 

First, we set $q=0$ and $p=1$, that means the bosonic part of target space is
just one copy of ${\mathcal M}_{Liouville} \times S^3 \times T^4$. 
The $p\neq 1$ case can be generated form the case $p=1$ just as in the 
bosonic string theory.

It is well known that the $N=4$ SCA admits a $so(4)=su(2)\times su(2)$ 
automorphism \cite{ss,yu}. In our case, the first $su(2)$ comes from
the $S_3$ part, the second $su(2)$ comes from one factor $su(2)$ of
the $so(4)$ acting on $T^4$.  We first decompose our $N=4$ SCA
into representations of the $su(2)\times su(2)$ automorphism.
The Virasoro generators, eq.(\ref{lfln}) belongs to $\bf{ (0,0)}$ and
the affine $su(2)$ generators, eq.(\ref{kn}) are in the $\bf{ (1,0)}$
The supercurrent $G_\al$'s belong to the $\bf{ (\hf,\hf)}$. To see the action
of the $su(2)\times su(2)$ on the $G_{\al}$'s, 
it is convenient to write $G_{\al}$'s in terms of real components $\Gt^A$'s
\bea
G_1&=&\Gt^0+i\Gt^3\\
G_2&=&\Gt^2+i\Gt^1\no
\eea
Define a matrix $\left(G^{AB}\right)$ to be 
\ben
G=i\Gt^A\sigma_A=\left(\begin{array}{rr}\Gt^0+i\Gt^3&\-\Gt^2+i\Gt^1\\
\Gt^2+i\Gt^1&\Gt^0-i\Gt^3\end{array}\right)=
\left(\begin{array}{rr}G_1&-\Gb_2\\
G_2&\Gb_1\end{array}\right)
\een
Here, $\sig^i, \ i=1,2,3$ are the Pauli matrices,
\ben
\sig^1=\left(\begin{array}{cc}0&1\\1&0\end{array}\right),\ \
\sig^2=\left(\begin{array}{cc}0&-i\\i&0\end{array}\right),\ \
\sig^3=\left(\begin{array}{cc}1&0\\0&-1\end{array}\right)
\een 
$\sig^0$ is proportional to identity matrix,
\ben
\sig^0=\left(\begin{array}{cc}i&0\\0&i\end{array}\right)
\een
and $\sig_A=(\sig^A)^*$.
Then it is clear that the first $su(2)$ acts on $G$ by left group action, and
the second $su(2)$ by right.
 
Now we also need to decompose the fields living in the transversal space
into the representations of the $su(2)\times su(2)$ automorphism. The
Liouville field $\p$ and its super partner (${\tps}^3$) from the \ads3 is in 
the $\bf{ (0,0)}$. The $K^A$'s and their super partner ${\chi}^A$'s 
are in the $\bf{ (1,0)}$. Finally, from $T^4$, we have the $u(1)$ currents 
$\pa X^N$'s and their super partner $\lambda^N$'s for $N=1,2,3$  in the 
$\bf{ (0,1)}$ and $\pa X^0$, $\lambda^0$ in the $\bf{ (0,0)}$. 
Since the
spacetime supercurrents $G_\al$'s are spacetime spinors, and the worldsheet 
fermions are spacetime vectors, we first need to
transform the later into spacetime spinors. This can be done by the usual
bosonization procedure.\\
Define
\bea
\chi^3+\sqrt{\frac{2}{k}}\tps^3&=&e^{\p^1}\no\\
\chi^3-\sqrt{\frac{2}{k}}\tps^3&=&e^{-\p^1}\no\\
\chi^+&=&e^{\p^2}\no\\
\chi^-&=&e^{-\p^2}\\
\lambda^3+i\la^0&=&e^{\p^3}\no\\
\lambda^3-i\la^0&=&e^{-\p^3}\no\\
\lambda^1+i\la^2&=&e^{\p^4}\no\\
\lambda^1-i\la^2&=&e^{-\p^4}\no
\eea
Then we can construct the $su(2)\times su(2)$ generators acting on these
fermions in term of $\p^A$'s
\bea
j^+&=&\oint e^{\p^2}(e^{\p^1}+e^{-\p^1})\no\\
j^-&=&\oint (e^{\p^1}+e^{-\p^1})e^{-\p^2}\no\\
j^3&=&\oint \pa \p^2\\
k^+&=&\oint e^{\p^4}(e^{\p^3}+e^{-\p^3})\no\\
k^-&=&\oint (e^{\p^3}+e^{-\p^3})e^{-\p^4}\no\\
k^3&=&\oint \pa \p^4\no
\eea
Using the $\p^A$ fields and considering the representations of the $j^a$'s 
and $k^a$'s,
we can construct a set of worldsheet fermion fields which are also
spacetime spinors
\bea
\ps_1&=&e^{\hf(\p_1+\p_2+\p_3+\p_4)}\label{ps}\\
\psb_1&=&e^{\hf(-\p_1-\p_2-\p_3-\p_4)}\no\\
\ps_2&=&e^{\hf(-\p_1-\p_2+\p_3+\p_4)}\no\\
\psb_2&=&e^{\hf(\p_1+\p_2-\p_3-\p_4)}\no\\
\chi_1&=&e^{\hf(-\p_1+\p_2+\p_3-\p_4)}\no\\
\chb_1&=&e^{\hf(\p_1-\p_2-\p_3+\p_4)}\no\\
\chi_2&=&e^{\hf(\p_1-\p_2+\p_3-\p_4)}\no\\
\chb_2&=&e^{\hf(-\p_1+\p_2-\p_3+\p_4)}\no
\eea

The OPE of ${\ps}_{\al},~\chi_{\al}$ are
\bea
{\ps}_{\al}(z){\psb}_{\be}(w)=\frac{{\del}_{\al \be}}{z-w}+ \cdots \cdots\\
{\chi}_{\al}(z){\chb}_{\be}(w)=\frac{{\del}_{\al \be}}{z-w}+ \cdots \cdots
\eea
and all others are regular.

For free fermions $\chi_\al$ in $\bf{ (\hf,\hf)}$, and $u(1)$ currents
in $\bf{ (0,0)}$ and $\bf{ (0,1)}$ from $T^4$ part, it is easy to construct 
a $N=4$ SCA with $c=6$, cf.\cite{yu}
\bea
\Gh_1&=&\chi_1 \pa (X^3+iX^0)-\chb_2 \pa (X^1+iX^2) \label{gh}\\
\Ghb_1&=&\chb_1 \pa (X^3-iX^0)+\chi_2 \pa (X^1-iX^2)\no\\
\Gh_2&=&\chi_2 \pa (X^3+iX^0)+\chb_1 \pa (X^1+iX^2)\no\\
\Ghb_2&=&\chb_2 \pa (X^3-iX^0)+\chi_1 \pa (X^1-iX^2)\no\\
\hat{K^A}&=&\hf \chi {\sig}^A \chb\no\\
\hat{T}&=& -\hf \pa X^N \pa X_N +\hf(\pa\chi_1\chb_1-\chi_1\pa\chb_1
+\pa\chi_2\chb_2-\chi_2\pa\chb_2)\no 
\eea
Now we are ready to construct another $N=4$ SCA, which are orthogonal to the 
one given in eqs.(\ref{gh}), with the bosonic fields coming from 
$AdS_3\times S_3$
and free fermions $\ps_\al$'s defined in eq.(\ref{ps}).  
Let us first consider
all the possibilities for constructing $G_1$ up to a normalization factor
\ben
G_1={\ps}_1k^3+A{\ps}_2k^+-B\pa{\ps}_1+C{\ps}_1{\psb}_2{\ps}_2
+D{\ps}_1 \pa\p 
\een 
Here, A,B,C,D are numeric coefficients to be determined.\\
$K^+(z)G_1(w)$ regular determines $A=1$. $G_1(z)G_1(w)$ regular
determines $C=1$, $D=(\frac{k'}{2})^{\half}$. Now consider the
OPE $K^-(z)G_1(w)$. By  demanding that only simple pole appear determines
$B=k'-1$.
Now $G_1$ is determined up to a normalization factor. By $su(2)\times su(2)$
automorphism, $G_2$ and $\Gb_1$, $\Gb_2$ can also be determined. Finally 
we have  
\bea
G_1&=&\sqrt{\frac{2}{k'}}[~{\ps}_1k^3+{\ps}_2k^+-(k'-1)\pa{\ps}_1
+{\ps}_1{\psb}_2{\ps}_2+(\frac{k'}{2})^{\half}{\ps}_1 \pa\p] \\
G_2&=&\sqrt{\frac{2}{k'}}[-{\ps}_2k^3+{\ps}_1k^- -(k'-1)\pa{\ps}_2
+{\ps}_2{\psb}_1{\ps}_1+(\frac{k'}{2})^{\half}{\ps}_2\pa\p] \no\\
\bar G_1&=&\sqrt{\frac{2}{k'}}[{\psb}_1k^3+{\psb}_2k^- +(k'-1)\pa{\psb}_1
+{\psb}_2{\ps}_2{\psb}_1-(\frac{k'}{2})^{\half}{\psb}_1\pa\p] \no\\
\bar G_2&=&\sqrt{\frac{2}{k'}}[-{\psb}_2k^3+{\psb}_1k^+ +(k'-1)\pa{\psb}_2
+{\psb}_1{\ps}_1{\psb}_2-(\frac{k'}{2})^{\half}{\psb}_2\pa\p] \no
\eea
It is easy to check
\bea
K^A(z)G_{\al}(w)&=&\frac{\half G_{\be}({\sig}^A)_{\be \al}}{z-w}+ 
\cdots \cdots\\
K^A(z)\bar G_{\al}(w)&=&-\frac{\half ({\sig}^A)_{\al \be} \bar G_{\be}}{z-w}+
\cdots \cdots \no
\eea
And
\bea
K^A(z)K^B(w)&=&\frac{\half(k'-1)}{(z-w)^2}+\frac{i{\eps}^{ABC}K^C}{z-w}
+ \cdots \cdots \\
G_{\al}(z)\bar G_{\be}(w)&=&\frac{2T}{z-w}+
\frac{2{\sig}_A\pa K^A}{z-w}+\frac{4{\sig}_B K^B}{(z-w)^2}
+\frac{4(k'-1)}{(z-w)^3}+ \cdots \cdots \no \\
T(z)G_\al(w)&=&\frac{G_\al}{z-w}+\frac{3/2~G_\al}{(z-w)^2}+\cdots\cdots\no
\eea
Here
\ben
T=\frac{1}{k'}k^Ak_A+\half(\pa{\ps}_1{\psb}_1-{\ps}_1\pa{\psb}_1
+\pa{\ps}_2{\psb}_2-{\ps}_2\pa{\psb}_2)-\half\pa\p\pa\p
+(\frac{1}{2k'})^{\half}(k'-1){\pa}^2\p\label{ft}
\een
The central charge of this spacetime Virasoro algebra is $c=6(k'-1)$.

Considering the direct sum of the two orthogonal $N=4$ SCAs constructed above,
we get our final $N=4$ SCA on the space $AdS_3\times S_3\times T^4$ with the 
central charge $c=6k'$, the desired result.

\bea
G^{tot}_{\al,n}&=&\sqrt{\frac{1}{p}}\oint dz 
(G_{\al}+\Gh_{\al})e^{qn}z^{pn+\hf}\label{gstn}\\
\bar G^{tot}_{\al,n}&=&\sqrt{\frac{1}{p}}\oint dz (\bar{G_{\al}}+\Ghb_{\al})
e^{qn}z^{pn+\hf}\\
K^{tot,A}_n&=&\oint dz (K^A+\hat{K}^A)e^{qn}z^{pn}\\
L^{tot}_n&=&\frac{1}{p}\oint dz [(T+\hat T)e^{qn}z^{pn+1}]
+\frac{c}{24}(p-\frac{1}{p})\delta_{n,0}\label{lstn}
\eea
where $c=6k'$. The $L_n$ should be equivalent to that in eq.(\ref{lfln}), 
then we get $k'=k$ 

\subsection{Orbifold $Sym_p{\mathcal M}/Z_p$ as the
Target Space of $N=4$ SCFT}\label{orb}

Superstring theory on \ads3 space defines a spacetime $N=4$ superconformal
field theory living on the 2d boundary of \ads3. 
Let us denote the target space of the spacetime 
superconformal field theory for $p\equiv \oint \g^{-1}\pa\g=1$ by 
$\mathcal{M}$. We shall show in this section that for general 
integer $p$, the target space is the orbifold $Sym_p{\mathcal{M}}/Z_p$, 
where, $Sym_p$ means symmetric product of $p$ objects,
and $Z_p$ is the cyclic group of order $p$. In the following  we sketch 
the proof.

Let us first consider the $N=4$ theory on the space $Sym_p\mathcal{M}$, which 
consists of $p$ copies of subspace $\mathcal{M}$, labeled by 
${\mathcal{M}}^i,\ i=1,2,\cdots,p$.
On each ${\mathcal{M}}^i$, we have an independent $N=4$ theory corresponding to $p=1$, 
which are constructed by the holomorphic fields living in the transversal space
in the light-cone gauge. In our case, those fields are 
$\p^i,~ K^{A,i},~\ps_\al^i,~\chi_\al^i,~\pa X^{N,i}$, which we denote 
collectively by ${\mathcal{J}}^{A,i}$. The generators of the $N=4$ SCA on 
${\mathcal{M}}^i$, such as 
$T^i,\ G_\al^i,\ K^{A,i}$ are collectively denoted by ${\mathcal T}^{A,i}$, with the
generators of the total $N=4$ SCA on $Sym_p\mathcal{M}$ given by
\ben
{\mathcal T}^A =\sum_i {\mathcal T}^{A,i}\label{ct}
\een
Now consider the  $N=4$ theory on $Sym_p{\mathcal{M}}/Z_p$. 
The $Z_p$ cyclic group is an Abelian group generated by a single element
$\hat{\la}, \ \hat{\la}^p=identity$.  
$\hat{\la}^j$ acts on the  ${\mathcal{J}}^{A,i}$ 
by shifting the $i$ index by $j$ units: 
${\mathcal{J}}^{A,i}\ra {\mathcal{J}}^{A,i+j}$. 
Gauging by the $Z_p$ action means that the 
${\mathcal{J}}^{A,i}$'s are defined up to $Z_p$ actions. That is to say, 
${\mathcal{J}}^{A,i}$'s admit a $Z_p$ hololomy around a non-contractable
loop in $z$ complex plane, which represents the \ads3 boundary.
For example, a possible choice of the boundary condition could be
\ben
{\mathcal{J}}^{A,i}(ze^{2\pi i})={\mathcal{J}}^{A,i+j}(z)\label{j}
\een
It turns out that any given choice of $j$ in eq.(\ref{j}) results in
an equivalent theory. Thus it suffices to consider the conjugacy class
of the $Z_p$ action, which can be represented by a single element 
$\hat{\la}$ corresponding to $j=1$ in eq.(\ref{j}).
It is convenient to diagonalize the $Z_p$ action in eq.(\ref{j}).
Define:
\bea
\la= e^{2\pi i/p}\\
K^{A,[j]}=\sum_{i=0}^{p-1}\la^{-ij} K^{A,i}\label{dia}\\
\pa \p^{[j]}=\sum_{i=0}^{p-1}\la^{-ij} \pa\p^i\no\\
\pa X^{N,[j]}=\sum_{i=0}^{p-1}\la^{-ij} \pa X^{N,i}\no\\
\ps_\al^{[j]}=\frac{1}{\sqrt{p}}\sum_{i=0}^{p-1}\la^{-ij} \ps_\al^i\no\\
\chi_\al^{[j]}=\frac{1}{\sqrt{p}}\sum_{i=0}^{p-1}\la^{-ij} \chi_\al^i\no
\eea
conversely we have
\bea
K^{A,i}=\frac{1}{p}\sum_{j=0}^{p-1}\la^{ij} K^{A,[j]}\\
\pa \p^j=\frac{1}{p}\sum_{i=0}^{p-1}\la^{ij} \pa\p^{[i]}\no\\
\pa X^{N,j}=\frac{1}{p}\sum_{i=0}^{p-1}\la^{ij} \pa X^{N,[i]}\no\\
\ps_\al^{j}=\frac{1}{\sqrt{p}}\sum_{i=0}^{p-1}\la^{ij} \ps_\al^{[i]}\no\\
\chi_\al^{j}=\frac{1}{\sqrt{p}}\sum_{i=0}^{p-1}\la^{ij} \chi_\al^{[i]}\no
\eea

From eq.(\ref{j}) by setting $j=1$, we can read off the mode expansions for 
${\mathcal{J}}^{A,[i]}(z)$
\bea
K^{A,[i]}(z)&=&\sum_{n\in \bf{Z}}K^{A,[i]}_{n-i/p}z^{-n-1+i/p}\label{mod}\\
\pa\p^{[i]}(z)&=&\sum_{n\in \bf{Z}}a^{[i]}_{n-i/p}z^{-n-1+i/p}\no\\
\pa X^{N,[i]}(z)&=&\sum_{n\in \bf{Z}}a^{N,[i]}_{n-i/p}z^{-n-1+i/p}\no\\
\ps_\al^{[i]}(z)&=&\sum_{n\in \bf{Z}}b^{[i]}_{\al,n-\hf-i/p}z^{-n+i/p}\no\\
\chi_\al^{[i]}(z)&=&\sum_{n\in \bf{Z}}c^{[i]}_{\al,n-\hf-i/p}z^{-n+i/p}\no
\eea

From the mode expansion in eq.(\ref{mod}), it can be seen that the 
normalizations in eq.(\ref{dia}) are chosen in such a way that
\bea
&&[K^{A,[i]}_{n-i/p}, K^{B,[j]}_{m-j/p}]=f^{ABC}K^{C,[i+j]}_{n+m-(i+j)/p}
+ \delta_{i+j,0}\delta_{n+m,0}\delta_{a,b}k(pn-i)\\
&&[b^{[i]}_{\al,n+\hf-i/p},\bar b^{[j]}_{\be,m-\hf-j/p}]=\delta_{i+j,0}\delta_{n+m,0}\no
\eea
And similar commutation relations for 
$a^{[i]}_{n-i/p},~ a^{N,[i]}_{n-i/p},~c^{[i]}_{\al,n-\hf-i/p}$. 
Those commutators are to be compared with
that of the $N=4$ theory with the target space 
$\mathcal{M}$, for which,
we have the mode expansions
\bea
K^A(z)&=&\sum_{n\in \bf{Z}}K^A_n z^{-n-1}\label{mod2}\\
\pa\p(z)&=&\sum_{n\in \bf{Z}}a_n z^{-n-1}\no\\
\pa X^N(z)&=&\sum_{n\in \bf{Z}}a^N_n z^{-n-1}\no\\
\ps_\al(z)&=&\sum_{n\in \bf{Z}}\ps_{\al,n-\hf}z^{-n}\no
\eea
Compare eqs.(\ref{mod2}) with eqs.(\ref{mod}), we are led to the following 
isomorphism:
\bea
K^{A,[i]}_{n-i/p}&\ra& K^A_{pn-i}\label{iso}\\
a^{[i]}_{n-i/p}&\ra& a_{pn-i}\no\\
a^{N,[i]}_{n-i/p}&\ra& a^N_{pn-i}\no\\
b^{[i]}_{\al,n-\hf-i/p}&\ra& b_{\al,np-\hf p-i}\no\\
c^{[i]}_{\al,n-\hf-i/p}&\ra& c_{\al,np-\hf p-i}\no
\eea
Depending on $p$ even or odd, for $N=4$ theory on  
$Sym_p{\mathcal{M}}/Z_p$ with
NS boundary condition, we need to define a $N=4$ theory on $\mathcal{M}$ with 
Ramand or NS boundary condition.

Now we are ready to express the generators for the total $N=4$ SCA on 
$Sym_p{\mathcal{M}}/Z_p$, eq.(\ref{ct}), in terms of the $Z_p$ twisted 
fields in eq.(\ref{mod}). Notice that when the fields ${\mathcal{J}}^{A,[i]}$
are twisted, the zero mode of the energy-momentum tensor acquires a shift,
similar to the anomalous transformation of the  energy-momentum tensor
under the conformal transformation $z\ra z^{\frac{1}{p}}$, cf. 
eq.(\ref{dlln2}).
For simplicity, let us consider the $T(z)$ in eq.(\ref{ft}) as an 
example.
\bea
&&T(z)\label{ft2}\\
&=&\sum_{i}T^{i}+\frac{c}{24}(p-\frac{1}{p})z^{-2}\no\\
&=&\sum_i \frac{1}{k'}k^{A,i}k_A^{i}
+\half(\pa{\ps}_\al^{i}{\psb}_\al^{i}-
{\ps}_\al^{i}\pa{\psb}_\al^{i}-\pa\p^{i}\pa\p^{i})
+(\frac{1}{2k'})^{\half}(k'-1){\pa}^2\p^{i} \no\\
&=&\sum_{i,j,l}\frac{ \la^{ij}\la^{il}}{p^2}(\frac{1}{k'}k^{A,[j]}k_A^{[l]}+
\half(\pa{\ps}_\al^{[j]}{\psb}_\al^{[l]}-
{\ps}_\al^{[j]}\pa{\psb}_\al^{[j]}-\pa\p^{[j]}\pa\p^{[j]})\no\\
&&+\sum_{i,j}\frac{\la^{ij}}{p}(\frac{1}{2k'})^{\half}(k'-1){\pa}^2\p^{[j]}\no\\
&=& \sum_{i}\frac{1}{p}\{\frac{1}{k'}k^{A,[i]}k_A^{[i]}+
\half(\pa{\ps}_\al^{[i]}{\psb}_\al^{[i]}-
{\ps}_\al^{[i]}\pa{\psb}_\al^{[i]}-\half\pa\p^{[i]}\pa\p^{[i]})\} 
+(\frac{1}{2k'})^{\half}(k'-1){\pa}^2\p^{[0]}\no
\eea 

Substituting eq.(\ref{iso}) into the last line of eq.(\ref{ft2}) and making
similar changes on $\hat T(z)$, we arrive at the eq.(\ref{lstn}).
 Similar relation holds for $G_\al(z),\ K^A(z)$. Thus the $N=4$ SCA defined in 
eqs.(\ref{gstn}-\ref{lstn}) 
in fact defines a theory on the target space $Sym_p{\mathcal{M}}/Z_p$.

\section{GS Superstring Theory on 
$AdS_3 \times S_3 \times T_4 $ Spacetime with NS-NS Background}

In the the section Introduction, we have briefly discussed the difficulty 
arising, when quantizing
the Green-Schwarz type superstring theory on $AdS_3 \times S_3 \times T_4 $, 
due to the high nonlinearality of the equations of motion. 
In this section, we shall
describe how to generalize the light-cone gauge choice to the case of GS string
theory on \ads3 space, in which we get linear equations of motion. In the 
present
paper, we shall only consider the case with pure NSNS background. The RR 
charged one is more complicated and deserves a separate 
consideration\cite{later}

The action for the GS string theory on the space $AdS_3 \times S_3 \times T_4 $
has been presented in refs.\cite{raj, pes, par}. Here, following ref.\cite{pes} and using 
the convention adopted in Appendix A, we
write down the $\kappa$ symmetry fixed GS superstring
action in pure NSNS background.

\bea
S=&&-\half \int d^2 \sig\{(\sqrt{g}{g}^{ij}
+{\eps}^{ij})r^2[{\pa}_ix^+
+\hf({\psb}_L{\pa}_i{\ps}_L-{\pa}_i{\psb}_L{\ps}_L)]\label{sns}\label{gsa}\\
&&~~~~~~~~~~~~~~~\times [{\pa}_jx^-+\hf({\psb}_R{\pa}_j
{\ps}_R-{\pa}_j{\psb}_R{\ps}_R)]\no\\
&&~~~~~~~+\sqrt{g}g^{ij}\frac{l^2}{l_s^2}[r^{-2}{\pa}_ir{\pa}_jr
+tr(\pa_i g\pa_j g^{-1})]
+{\del}_{rs}{\pa}_iz^r{\pa}_jz^s\}\no\\
&&+\frac{i}{3}\int_{M^3}d^3x\frac{l^2}{l_s^2}\eps^{ijk}
tr(g^{-1}\pa_igg^{-1}\pa_jgg^{-1}\pa_kg)\no
\eea
where, $z^r$ are the bosonic fields 
on $T^4$, and $g$ is a group element of $SU(2)$. 
Using the projective coordinate of $S^3$ (the group manifold of $SU(2)$), 
eq.(\ref{gsa}) becomes the same action in ref.\cite{pes}.

In ref.\cite{raj}, the final form of the action involves only pure RR background
and the spinors do not conform to our convention.
In the following we shall describe how the $\kappa$ symmetry is fixed 
leading to the same action as in eq.(\ref{gsa}), in the case when pure NSNS
background is considered.  

In particular, 
we shall show here, that the gauge choice in ref.\cite{raj} 
\ben
{\theta}_{R-}^I=\half({\del}^{IJ}-i{\eps}^{IJ}{\g}_R^0{\g}_R^1){\theta}_R^J=0
\label{thr}
\een
is just the light-cone gauge choice for the spacetime spinors. 
Here the lower index R refers to the gamma matrices and spinors in 
ref.\cite{raj}.

Now we rewrite eq.(\ref{thr}) in our convention
\bea
{\theta}_-^I&=&T^{-1}{\theta}_{R-}^I\\
&=&\half({\del}^{IJ}-i{\eps}^{IJ}T^{-1}{\g}^0_R{\g}^1_RT){\theta}^J\no\\
&=&0\no
\eea
i.e.
\ben
{\theta}_1^I+i{\theta}^{II}_1=0\qquad {\theta}_2^I-i{\theta}^{II}_2=0 
\label{th}
\een
And
\ben
{\bar{\theta}}_2^I-i{\bar{\theta}}^{II}_2=0\qquad {\bar{\theta}}_1^I \label{thb}
+i{\bar{\theta}}^{II}_1=0
\een
Set
\bea
&&{\ps}_{R\al}=\half({\theta}_{\al}^I-i{\theta}_{\al}^{II})\qquad
{\ps}_{L\al}=\half({\theta}_{\al}^I+i{\theta}_{\al}^{II})\\
&&{\psb}_{R\al}=\frac{i}{2}({\bar {\theta}}_{\al}^I+i 
\bar {\theta}_{\al}^{II})\qquad
{\psb}_{L\al}=-\frac{i}{2}({\bar {\theta}}_{\al}^I-i 
\bar {\theta}_{\al}^{II})\no
\eea
Then eqs.(\ref{th}-\ref{thb}) lead to
\ben
\Gamma^+\ps_{R}=\Gamma^+\psb_{L}=\Gamma^-\ps_{L}=\Gamma^-\psb_{R}=0\label{lcg}
\een
where $\Gamma^+=\Gamma^1+\Gamma^2$ and $\Gamma^-=\Gamma^1-\Gamma^2$. 

Here we ignore the indices of the spinor representation of $su(2)$.
We realize that eq.(\ref{lcg}) is the same as fixing the $\kappa$ symmetry 
in the light-cone gauge for the spacetime spinors in the flat spacetime, 
cf. \cite{gsw}.

Using $\ps_{R},~\ps_{L},~\psb_{R},~\psb_{L}$ to denote the nonzero
component of the spinors, then the action in ref.\cite{raj} with pure 
NSNS background can be written in the same form as given in ref.\cite{pes}. 

To compare eq.(\ref{sns}) with the NSR string, we introduce an auxiliary field $\be$
as the  Lagrange multiplier. The action, eq.(\ref{gsa}), then  becomes
\bea
S=&&\hf \int dzd \bar z \{ -\be\bar{\be}exp(-\frac{2}{\alp}\p)
+\be[\bar{\pa}x^++\hf({\psb}_L\bar{\pa}{\ps}_L-\bar{\pa}{\psb}_L{\ps}_L)]\\
&&+\bar{\be}[{\pa}x^-+\hf({\psb}_R{\pa}{\ps}_R-{\pa}{\psb}_R{\ps}_R)]
+\pa\p\pab\p-\frac{2}{\alp}\hat R^{(2)}\p\}\no\\
&&-\hf\int d^2\sig\sqrt{g}g^{ij}[\frac{l^2}{l_s^2}tr(\pa_i g\pa_j g^{-1})
+\pa_iz^r\pa_jz^r\no]\\
&&+\frac{i}{3}\int_{M^3}d^3x\frac{l^2}{l_s^2}\eps^{ijk}
tr(g^{-1}\pa_igg^{-1}\pa_jgg^{-1}\pa_kg)\no
\eea
where $\p=\alp log~r$ and $\alp^2=2k=2\frac{l^2}{l_s^2}$. 

The $S^3$ part of 
this action is just a WZNW model in $SU(2)$ group manifold. 
Because the radiuses 
of $AdS_3$ and $S^3$ are the same, it is obvious that the $sl(2,R)$ and $su(2)$ 
currents have the same level, as expected from subsection \ref{4.1}.

Regarding $\be\beb exp(\frac{-2\p}{\alp})$ as a screen charge, we get the 
the following linearized equations of motion for free bosons and fermions.
\bea
&&\pab \be=0 \qquad \pa \bar{\be}=0\no\\
&&\pab {\ps}_L=0 \qquad \pa {\ps}_R=0\\
&&\pab \bar{\ps}_L=0\qquad \pa \bar{\ps}_R=0\no\\
&&\pab x^+=0 \qquad \pa x^-=0\no
\eea
It is clear now that the light-cone gauge choice for the spinors in the 
GS type action is self consistent, and we can quantize the action by 
introducing OPE of free fields.
Fixing the $\kappa$ symmetry is equivalent to fixing the redundant fermionic 
degrees of freedom, just as what we have done in 
the light-cone
gauge quantization of the NSR string. Now the bosonic part of this 
action is the  same as NSR formulation. So, in $AdS_3\times S^3\times T^4$ 
spacetime, GS string in pure NSNS field background is equivalent to NSR 
superstring,
as can be seen by taking the light-cone gauge.
The rest part of quantization procedure goes exactly the same way as the
light-cone gauge quantization in the NSR formalism.

\section{Conclusions and Discussions}
In conclusion, we have presented the light-cone gauge quantization procedure 
for strings propagating
on \ads3 space charged with NS-NS fields. The procedure works in that
we get the right spacetime conformal anomalies for critical bosonic string 
as well as 
superstrings. The $N=4$ spacetime superconformal algebra arises in a 
natural way for
light-cone gauge fixed superstrings, both in NSR and GS formalism. 
Further, we have proven that in our light-cone gauge choice, the
resulting theory is unitary, hence a well define quantum theory.

To quantize superstring theory charged with R-R fields, it seems more
appropriate to formulate the string theory in a way with manifest
spacetime supersymmetries, as what has been done in refs.\cite{raj,pes}. 
However,
the classical equations of motion derived this way are highly nonlinear,
posing difficulties in choosing a convenient quantization procedure.
The same non-linearity problem in the case of flat spacetime is solved
by choosing a light-cone gauge. We believe that in the case of conformal
invariant background such as $AdS_3 \times S_3 \times T^4$ or $AdS_5 \times
S_5$, a generalized light-cone gauge choice can always be found.
The case with NS-NS background worked out in this paper provide just
an example, and such quantization procedure should be generalizable
to the cases involving R-R charges. Currently, we are making progress along
such directions.\cite{later}

In the present paper, we are assuming that the worldsheet theory corresponding
to the \ads3 space is a standard WZNW theory. However for $p> 0$, the 
holomorphic conformal scalar field $\g$ vanishes at the worldsheet origin, 
and $\be$, which has conformal weight one, is singular at the origin. 
In fact, $\g(z)\sim z^p, \ \be(z)\sim z^{-p-1}$ when $z\ra 0$. 
This is equivalent to the 
shifting of the Bose sea level, and hence shifting  the vacuum state in
the WZNW theory, cf. \cite{fms}. That would has some effects when coming to 
the calculations 
of the correlation functions on the worldsheet. Such point is not elaborated 
in our present paper and will be left for future discussions. 

\noindent{\bf Note Added:} After the completion of the present work, we became aware of
the paper, ref.\cite{sug}, the authors of which found  essentially the 
same target 
space for the $N=4$ SCFT as what we presented in Sec.4.2 from a different 
approach.

\section*{Acknowledgment}
It is pleasant to acknowledge our discussions with M. Li, Y. Gao and C. Zhu.

This work is supported in part by grants from the NSFC.

\section*{Appendix A}
The $N=2$ SUSY algebra $su(1,1|2)^2$, cf. ref.\cite{raj}, can be presented as
\bea
&&\{ Q_I, \bar Q_J \}=2{\del}_{IJ}(iP_a{\g}^A-P_{a'}{\g}^{a'})+{\eps}_{IJ}
(J_{ab}{\g}^{ab}-J_{a'b'}{\g}^{a'b'}) \no\\
&&[P_a, Q_I]=-\frac{i}{2}{\eps}_{IJ}{\g}_aQ_J \qquad
[P_{a'},Q_I]=\frac{1}{2}{\eps}_{IJ}{\g}_{a'}Q_J \no\\
&&[J_{ab}, Q_I]=-\half{\g}_{ab}Q_I \qquad
[J_{a'b'}, Q_I]=-\half{\g}_{a'b'}Q_I \\
&&[P_a, \bar Q_I]=\frac{i}{2}\bar Q_J{\eps}_{JI}{\g}_a \qquad
[P_{a'},\bar{Q}_I]=-\half \bar Q_J{\eps}_{JI}{\g}_{a'} \no\\
&&[J_{ab}, \bar Q_I]=\half\bar Q_J{\g}_{ab} \qquad
[J_{a'b'},\bar{Q}_I]=\half\bar Q_J{\g}_{a'b'}\no\\ 
&&[J_{AB},J_{CD}]={\eta}_{BC}J_{AD}+{\eta}_{AD}J_{BC}-{\eta}_{AC}J_{BD}
-{\eta}_{BD}J_{AC}\no\\
&&[J_{A'B'},J_{C'D'}]={\del}_{B'C'}J_{A'D'}+{\del}_{A'D'}J_{B'C'}
-{\del}_{A'C'}J_{B'D'}-{\del}_{B'D'}J_{A'C'}\no\\
&&P_a=J_{0a},\qquad P_{a'}=J_{0a'}\no
\eea
where, $Q_{I\al\al'}$ are the 2 complex chiral 6D spinors, 
with $I=I,II,~\al=1,2,~\al'=1,2$. 
The convention we use here is
\bea
&&A,B,C,D=-1,1,2,3 \qquad {\eta}_{AB}=(-1,-1,1,1)\no\\
&&A',B',C',D'=0,4,5,6 \qquad {\eta}_{A'B'}=(1,1,1,1)\no\\
&&a,b...=1,2,3 \qquad {\eta}_{ab}=(-1,1,1)\\
&&a',b'...=4,5,6 \qquad {\eta}_{a'b'}=(1,1,1)\no\\
&&{\g}^3=-{\sig}^3 \qquad {\g}^1=i{\sig}^2 \qquad {\g}^2={\sig}^1\no\\
&&{\g}^4={\sig}^1 \qquad {\g}^5={\sig}^2 \qquad {\g}^6={\sig}^3\no
\eea
In the following we use $\g^a,~\g^{a'}$ to represent  
$\g^a \otimes 1_2,~1_2\otimes \g^{a'}$ resp.

The Dirac matrices $\g^a$ used here is a little different from that in 
ref.\cite{raj}.
That means we use another basis of the spinors. If we use $Q_R$ representing
the SUSY 
charge in \cite{raj}, the relation between these two conventions are
\bea
&&Q=TQ_R\qquad \theta=T^{-1}{\theta}_R\\
&&{\g}^3=-T{\g}^1_RT^{-1}\qquad {\g}^1=T{\g}^3_RT^{-1}
\qquad {\g}^2=T{\g}^2_RT^{-1}\no
\eea
where, $T$ is a unitary transformation of the spinors.
\ben
T=\sqrt{\half} \left( \begin{array}{ccc}-i&-i&\\1&-1&\end{array}\right)
\een
From now on we separate the bosonic part algebra $so(2,2)\times so(4)$ as 
$(so(1,2)\times so(3))_L \times (so(1,2) \times so(3))_R$, and just concentrate
ourselves on the left-moving part only.

The algebra can be written as
\bea
J^1&=&-\frac{i}{2}(J_{23}-J_{-11}) \qquad K^4=-\frac{i}{2}(J_{56}+J_{04})\no\\
J^2&=&-\frac{i}{2}(J_{31}+J_{-12}) \qquad K^5=-\frac{i}{2}(J_{64}+J_{05})\\
J^3&=&-\frac{i}{2}(J_{12}+J_{-13}) \qquad K^6=-\frac{i}{2}(J_{45}+J_{06})\no\\
\eea
Actually, $so(1,2)$ is equivalent to $sl(2,R)$, so we can write the algebra
as $L_0, L_1, L_{-1}$.
\ben
L_0=iJ^3\qquad L_{\pm 1}=i(J^1\pm J^2)
\een

The super charges can also be combined to form a left-moving one, 
\bea
Q_{\al\al'}&=&\hf(Q_{I\al\al'}+iQ_{II\al\al'})\no\\
\bar Q_{\al\al'}&=&\hf(i\bar Q_{I\al\al'}+\bar Q_{II\al\al'})
\eea
It is easy to check
\bea
&&[J^a,~Q_{\al\al'}]=\frac{i}{2}\g^aQ_{\al\al'} \qquad 
[J^a,~\bar Q_{\al\al'}]=-\frac{i}{2}\bar Q_{\al\al'} \g^a\\
&&[K^{a'},~Q_{\al\al'}]=-\frac{1}{2}\g^aQ_{\al\al'} \qquad 
[K^{a'},~\bar Q]=\frac{1}{2}\bar Q_{\al\al'} \g^a\no
\eea
Set
\bea
G_{\hf\al'}&=&\bar Q_{1\al'}\qquad \bar G_{-\hf\al'}=Q_{1\al'}\\
G_{-\hf\al'}&=&\bar Q_{2\al'}\qquad \bar G_{\hf\al'}=-Q_{2\al'}\no
\eea
$L_0,L_1,L_{-1},K^{a'}$ and $G, \bar G$ close an algebra
\bea
&&[L_0,L_{\pm 1}]=\mp L_{\pm 1}\qquad [L_{+1},L_{-1}]=2L_0\\
&&[L_n,G_r]=(\half n -r)G_{n+m}\no\\
&&[G_r,\bar G_s]=2L_{r+s}+2(r-s){\sig}_{a'}J^{a'}_{r+s}\no
\eea
which is exactly the left-moving part of the global $N=4$ 
SCA in NS sector.

Extending this 6D super algebra to 10D is straightforward. 
The $(9+1)$D gamma matrices can be expressed using $\g^a,\g^{a'},\g^r$.
\bea
\Gamma^a&=&\g^a\otimes(1_2\otimes 1_4)\otimes \sig^1\\
\Gamma^{a'}&=&1_2\otimes(\g^{a'}\otimes \bar \g^5)\otimes \sig^2\no\\
\Gamma^r&=&1_2\otimes(1_2\otimes \g^r)\otimes \sig^2\no
\eea
where $\g^r$ (r=7,8,9,10) are the 4D gamma matrices, and 
$\bar \g^5=\g^7\g^8\g^9\g^{10}$.
And the preceding discussions on the relation between the superalgebra 
$su(1,1|2)^2$ and the global part of the $N=4$ SCA go over to the 10D case.

\end{document}